\documentclass[aps,prl,twocolumn,showpacs,superscriptaddress,groupedaddress]{revtex4}  
\usepackage{amssymb,amsmath}

\usepackage{graphicx}
\usepackage{epstopdf}

\usepackage{epsfig}
\usepackage{dcolumn}   
\usepackage{bm}        
\usepackage{amssymb}   

\newcounter{fig}

\newcommand{\beq}{\begin{equation}}
\newcommand{\eeq}{\end{equation}}
\newcommand{\bea}{\begin{eqnarray}}
\newcommand{\eea}{\end{eqnarray}}

\usepackage{amsmath}

\begin{document}



\title{Topological inflation with graceful exit}

\author{Anja Marunovi\'c$^*$}
\author{Tomislav Prokopec\footnote{a.marunovic@uu.nl, t.prokopec@uu.nl}}

\affiliation{Institute for Theoretical Physics, Spinoza Institute
and Center for Extreme Matter and Emergent Phenomena,
 Utrecht University, Postbus 80.195, 3508 TD Utrecht, The Netherlands}



\begin{abstract}
\noindent
We investigate a class of models of topological inflation in which a
super-Hubble-sized global monopole seeds inflation. These models are
attractive since inflation starts from rather generic initial
conditions, but their not so attractive feature is that, unless
symmetry is again restored, inflation never ends. In this work we
show that, in presence of another nonminimally coupled scalar field,
that is both quadratically and quartically coupled to the Ricci
scalar, inflation naturally ends, representing an elegant solution
to the graceful exit problem of topological inflation. While the
monopole core grows during inflation, the growth stops after
inflation, such that the monopole eventually enters the Hubble
radius, and shrinks to its Minkowski space size, rendering it
immaterial for the subsequent Universe's dynamics. Furthermore, we
find that our model can produce cosmological perturbations that
source CMB temperature fluctuations and seed large scale structure
statistically consistent (within one standard deviation) with all
available data. In particular, for small and (in our convention)
negative nonminimal couplings, the scalar spectral index can be as
large as $n_s\simeq 0.955$, which is about one standard deviation
lower than the central value quoted by the most recent Planck
Collaboration.

\end{abstract}



\maketitle

\section{Introduction}
\label{Introduction}

Inflationary paradigm~\cite{Guth:1980zm,Starobinsky:1980te}
is currently the most successful model of
the very early Universe that evolves into a late time universe consistent
with all current astronomical observations~\cite{Ade:2015lrj}.
Most of inflationary models are driven by a potential energy of some scalar field,
and cosmological perturbations, that serve as seeds for large scalar structure,
are generated by amplified quantum fluctuation of a scalar field
(the so-called inflaton) during inflation.

Many aspects of inflation
are a blessing: once it starts, if it lasts long enough, it
solves the homogeneity, isotropy, causality and flatness problems,
and, as a bonus, it provides a natural mechanism for
the seed perturbations~\cite{Starobinsky:1979ty,Mukhanov:1981xt}
that well explain the observed properties of both the cosmic microwave
background (CMB) radiation and the Universe's large scale structure (LSS).

 Historically, already the early scalar models of inflation suffered from the
graceful exit problem~\cite{Guth:1980zm}, {\it i.e.} in those
theories, once started, inflation never ends. This is why
"new"~\cite{Linde:1981mu,Hawking:1981fz,Albrecht:1982wi} and
"chaotic"~\cite{Linde:1983gd} models of inflation were designed to
provide a successful exit from inflation, thereby solving the
graceful exit problem. A closer look at these solutions reveals that
all of these models suffer from a severe fine tuning of one sort or
the other: either the initial field ought to be finely tuned to be
very close to zero in a sizeable (super-Hubble) volume of space, or
the potential energy around the (true or local) minimum of the
potential at which inflation ends ought to be finely tuned to zero
to many digits. This latter problem is typically ignored by
inflationary practitioners, the argument being that this problem is
indistinguishable from the cosmological constant problem and hence
-- as long as we do not have a good solution to the cosmological
constant problem -- we do not have to worry about the fine tuning
associated with the end of inflation. Modern inflationary
models~\cite{Baumann:2009ds,Lyth:1998xn,Polchinski:2006gy} are no
exception: they suffer from one or more of fine tuning problems, the
principal ones being: initial conditions, graceful exit problem,
the tuning of the potential energy at the end of inflation (which is in disguise the cosmological constant problem),
and the choice of parameters in the model (that are {\it e.g.}
unnaturally small). Some of those problems are absent in the original
Starobinsky's model~\cite{Starobinsky:1979ty}, the Tsamis-Woodard
model (see for example~\cite{Tsamis:2014kda} and references
therein)~\footnote{In this model the cosmological constant is driven
to zero by the (quantum) backreaction of gravitons that are produced
during inflation. However, the validity of the model has not been
rigorously established. Currently, the best calculation is from the
distant 1996~\cite{Tsamis:1996qq}, where the authors have performed
a two-loop perturbative calculation of the stress-energy tensor and
removed the divergences by using a momentum cutoff regularization
that breaks the symmetries of the underlying space, and its results
are hence not reliable.}, and a recent model~\cite{Glavan:2015aqa}
in which the decay of the (measured) cosmological constant in a
model with a nonminimally coupled scalar~\cite{Glavan:2015ora} and
cosmic inflation are intimately related.

 Here we construct a topological inflation model which does not suffer from fine tuning problems,
in the sense that inflation starts from generic initial conditions and
ends naturally. This is, of course, true
provided one accepts that the phase transition scale of the Grand Unified Theory (GUT) at which
global monopoles form, is a natural scale involving no fine tuning of model parameters.

The paper is organized as follows. In the following section we
introduce the model. Next, we discuss initial conditions.
That is followed by a discussion of our main results. An important section is devoted to
a discussion of graceful exit, {\it i.e.} how inflation ends, but no detailed discussion of
preheating is presented. In the final section we conclude.


\section{Topological Inflation}
\label{Topological Inflation} \vskip -0.1cm

 Global monopoles are topological defects generically created at a phase transition
 by the Kibble mechanism~\cite{Kibble:1976sj} (at least of the order
  one per Hubble volume) if the effective field
mass matrix changes from having all positive eigenvalues to at least
one negative eigenvalue~\cite{Prokopec:2011ms,Lazzari:2013boa}.
 The (classical, bare)
action that governs the dynamics of global monopoles is
\beq S_\phi=\int d^4 x \sqrt{-g}\left(
            - \frac 12 g^{\mu\nu}(\partial_\mu\phi^a)(\partial_\nu\phi^a)
            - V(\phi^a)
                       \right),
\label{Monopole Action}
\eeq
%
with a Higgs-like of $O(3)$ symmetric potential
%
\beq
V(\phi^a)=\frac{\lambda}{4}\left(\phi^a\phi^a-\phi_0^2\right)^2\,,
\label{potentialA} \eeq
%
%
where $\lambda$ is a self-coupling and repeated indices $a$ indicate
a summation over $a=1,2,3$. The scalar field $\vec \phi=(\phi^a)$
($a=1,2,3$) consists of 3 real components. When $\phi^a\phi^a\equiv
\phi_0^2$, the vacuum exhibits a field condensate and the $O(3)$
symmetry of the action is spontaneously broken to an $O(2)$. The
vacuum manifold of the theory is the quotient space, $O(3)/O(2)$,
which is homeomorphic to the two-sphere, $S^2$. The
potential~(\ref{potentialA}) is chosen such that the energy density
of the (classical) vacuum is $V(\phi_0)=0$. One can view this
condition as fine tuning just as in any inflationary model. If
$V(\phi_0)>0$, there will be a residual positive cosmological
constant that outside the monopole core that drives eternal
inflation. However, in Ref.~\cite{Glavan:2015aqa}  was shown that,
when a suitably nonminimally coupled scalar field is added,
inflation can end also in this model.

 Let us for simplicity consider firstly the Einstein-Hilbert (EH) action for gravity,
%
\beq S_{EH}=\frac{1}{16\pi G_N}\int d^4 x \sqrt{-g}R \,, \label{E-H
action} \eeq
%
where $G_N$ is the Newton constant, $R$ is the Ricci curvature
scalar, $g$ is the determinant of the metric tensor $g_{\mu\nu}$.
The simplest topologically non-trivial solution of~(\ref{Monopole Action}--\ref{E-H action})
is a hedgehog-like spherically symmetric solution of the form,
\beq
\vec \phi(t,\vec r\,)
  = \phi(r)\left(\sin\theta\,\cos\varphi,\sin\theta\,\sin\varphi,
      \cos\theta\right)^T
\,,
\label{AnsatzSca}
\eeq
%
%
where $\theta, \varphi$ and $r$ are spherical coordinates. This
solution represents a global monopole -- it has a non-vanishing
vacuum expectation value which does not decay as it is stabilized by
topology (for more details see {\it e.g.}~\cite{Marunovic:2014hla}). This
is, however, the case for static monopoles for which the vacuum
expectation value is smaller than the reduced Planck mass, $\phi_0 \ll
M_{\rm P}$, $M_{\rm P}=\big(8\pi G_N\big)^{-1/2}$,
 or equivalently $\Delta\ll 1$, where $\Delta=8\pi G_N\phi_0^2$ is the
solid deficit angle (on the detailed analyses of static global
monopoles see {\it e.g.}~\cite{Marunovic:2013eka}).
On the other hand, if the vacuum expectation value is larger than
the reduced Planck mass, $\phi_0\gtrsim M_{\rm P}$, the monopole
becomes dynamical, {\it i.e.} it backreacts on the background space
such that it starts to inflate~\cite{Vilenkin:1994pv}. To be more
precise, detailed numerical analyses have shown that a topological
defect undergoes an inflationary expansion already for
$\phi_0\gtrsim 0.33 M_{\rm P}$ ($\Delta\gtrsim
0.1$)~\cite{Sakai:1995nh}.

Let us now estimate how the condition $\phi_0\gtrsim M_{\rm P}$ is translated
to the relation between the monopole size and the Hubble radius. The
monopole core size in a flat space-time is defined by the balance
between the gradient and potential energy,
$(\phi_0/\delta_0)^2\sim V(0)$, which for~(\ref{potentialA}) is
\beq \delta_0\simeq
\frac{2}{\sqrt{\lambda}\phi_0}=\frac{2}{\mu_0}\,,
\label{MonSize}\eeq
where $-\mu_0^2=\partial^2 V(0)/\partial\phi^2$ defines the curvature of the potential
at the origin, $\phi^a=0$.
If the monopole potential energy dominates over its kinetic energy
(which is the case in slow roll) and also over  energy densities of
other fields, then the Hubble parameter $H(t)$ is well approximated by the
following Friedmann
equation,
\beq H^2=\frac{8\pi G_N}{3}\, V(\phi)
\,,
\label{FrEq2}
\eeq
where $H=\dot a/a$ and the metric is
\beq
ds^2=-dt^2+a^2(t)d\vec r^{\,2}
\,.
\label{FrMetric}
\eeq
During inflation, the Hubble parameter changes adiabatically in time,
$|\dot H|\ll H^2$. In fact, in topological inflation one can approximate $V(\phi)\simeq V(0)$,
such that
\beq H_0^{-1}=M_{\rm P}\sqrt{\frac{3}{V(0)}}\simeq \frac{\sqrt{12} M_{\rm P}}{\sqrt{\lambda}\phi_0^2}
\,.
\label{HorizonSize}
\eeq
From Eqs.~(\ref{MonSize}) and~(\ref{HorizonSize}), it follows that
the condition  $\phi_0 \gtrsim M_{\rm P}$ leads to $\delta_0 \gtrsim H_0^{-1}$.

Topological inflation was originally investigated independently by
Vilenkin~\cite{Vilenkin:1994pv} and Linde~\cite{Linde:1994hy}. They
showed that, if the size of the defect is much smaller than the
Hubble radius, $\delta_0\ll H^{-1}$ ($\phi_0\ll M_{\rm P}$), gravity
does not considerably affect the monopole structure. On the
contrary, if the monopole size is much larger than the Hubble
radius, $\delta_0 >> H^{-1}$ ($\phi_0 >> M_{\rm P}$), the monopole
will drive inflation and moreover its core will grow during
inflation. This {\bf topological inflation}, once started, never
ends, {\it i.e.} it is {\it eternal}. An important advantage of
those kind of models is that inflation begins from generic initial
conditions, thereby removing one of the major fine tuning problems
of scalar inflationary models. In the next section we discuss in
some detail how generic initial conditions are that lead to
topological inflation. But before we do that, we address arguably
the most important unsolved problem of topological inflation: how to
exit from inflation. Parenthetically, we remark that in Ref.~\cite{ChoVilenkin} 
the authors analyzed
the spacetime structure of an inflating global monopole and showed
that there is no graceful exit problem. Even though this is true if
the monopole size is comparable to the Hubble radius, the gradient
terms in the potential will generate an inhomogeneous field and lead
to an anisotropic, inhomogenous expansion. Albeit 
we can say that the amount of inhomogeneities depends 
on the observer's position with respect to the monopole center: observers closer to
(further from) the center will in general observe less (more) inhohomogeneities,
the precise amount of inhomogeneities is not known, and this question therefore deserves 
to be investigated.

\subsection{Graceful exit}
\label{Graceful exit}

In order to ensure exit from inflation everywhere in space, we introduce another
scalar field $\psi$ that nonminimally coupls to gravity whose action is,
\bea S_\psi=\int d^4 x\sqrt{-g}\left(\frac 12 R F(\psi)-\frac 12 (\partial_\mu\psi)(\partial_\nu\psi)g^{\mu\nu}\right)\,,
\label{ActionPsi}\\
F(\psi)=M_{\rm P}^2-\xi_2\psi^2-\frac{\xi_4}{M_{\rm P}^2}\psi^4+{\cal O}(\psi^6)
\,, \qquad\;
\label{F Psi}
\eea
where $\xi_2$ and $\xi_4$ are dimensionless parameters (to be
determined from the recent Planck satellite measurements).
As we shall see, for sufficiently large monopole the field $\psi$ is the inflaton field.
In this work we assume that non-minimal couplings of the type, $\int d^4x\sqrt{-g}G(\vec \phi^2)R$,
are absent, and that $\psi$ and $\phi^a$ interact only indirectly via gravity. If would be of interest to 
investigate the effect of these couplings on predictions of the model.

In this paper we consider only homogeneous case (and neglect
gradient terms), which is justified when the size of the monopole is
much larger than the Hubble radius. Since the monopole
core grows during inflation, this approximation
is well justified if inflation lasts for much longer than
the required minimum, $N_0\gg 65$. In this case, the inflationary patch
that corresponds to today's observable universe takes
place close enough to the center of the monopole, that it can be
well approximated the potential~(\ref{potentialA}) at the core center,
\beq V(\phi)\approx V(0)=\frac{\lambda \phi_0^4}{4}\,. \label{V(0)}
\eeq

In our recent paper~\cite{Glavan:2015aqa} we have analyzed inflation driven by
a cosmological constant and nonminimally coupled scalar field with the
same action as Eqs.~(\ref{ActionPsi}--\ref{F Psi}). Here, we also
work in slow roll approximation, which is to a certain extent tested
in~\cite{Glavan:2015aqa}. In order to study inflation in slow roll regime, it
is more convenient to transform the full action to the Einstein
frame, with the canonically coupled scalar fields,
\bea g_{E\mu\nu}&=&\frac{F(\psi)}{M_{\rm P}^2}g_{\mu\nu}\,,\\
d\psi_E&=&\frac{M_{\rm P}}{F(\psi)}\sqrt{F(\psi)+\frac
32\left(\frac{dF(\psi)}{d\psi}\right)^2}d\psi\,. \eea
\bea S_E&=&\int d^4x\sqrt{-g_E}\bigg[\frac 12 M_{\rm P}^2 R_E -\frac 12
(\partial_\mu\phi_E)(\partial_\nu\phi_E)g_E^{\mu\nu}
\nonumber\\
&-&\frac 12 (\partial_\mu\psi_E)(\partial_\nu\psi_E)g_E^{\mu\nu}
     -V_E(\phi_E,\psi_E)\bigg]
\,,
\eea
where
\beq
V_E(\phi_E,\psi_E)=\frac{\lambda}{4}\left(\phi_E^2-\frac{\phi_0^2}{F(\psi)/M_{\rm P}^2}\right)^2\,,
\label{Potential VE}
 \eeq
and the frame transformation of $\phi_E$ is trivial,
\beq
\phi_E^2=\frac{\phi^2}{F(\psi)/M_{\rm P}^2}
\,.
\eeq
Since in this paper we investigate only the homogeneous case ($\delta_0\gg
H_0^{-1}$), the potential~(\ref{Potential VE}) can be approximated,
\beq
V_E(\phi_E,\psi_E)\simeq V_E(\psi)\simeq \frac{\lambda\phi_0^4
M_{\rm P}^4}{4F(\psi)^2}
\,,
\label{Potential VE Hom}
\eeq
At early stages of inflation when $\psi\ll M_{\rm P}$, our model mimics
"hilltop inflation"~\cite{Lyth:1998xn,Boubekeur:2005zm}),
but at late times a better approximation is power law inflation (driven by an exponential potential),
see Ref.~\cite{Glavan:2015aqa}.
The potential~(\ref{Potential VE Hom}) describes a one-field inflationary model in which $\psi$ is the inflaton 
and it approximates well the true two-field dynamics when the curvature of the potential in the $\psi$ direction is 
much larger than in the $\phi$ direction, {\it i.e.} when
 $-\mu_\psi^2 = -\xi_2\lambda\phi_0^4/M_P^2\gg -\mu_\phi^2 = \lambda\phi_0^2$. This is satisfied when 
the mopole size, $\phi_0\gg M_{\rm P}/\sqrt{-\xi_2}$, which we assume to hold true.

\begin{figure}
\centering
\hskip -0.2cm
\includegraphics[scale=0.7]{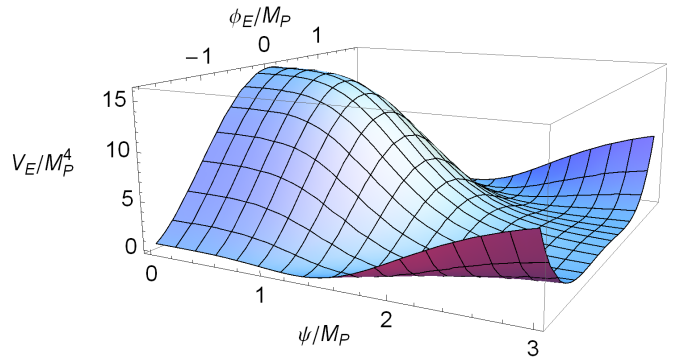}
\caption{Potential in the Einstein frame $V_E(\phi_E,\psi_E)$ for
$\xi_2=-0.002$ and $\xi_4=-0.1$. For a very large monopole
(homogeneous case), inflation takes place at $\phi_E\simeq 0$.
Inflaton field $\psi_E\simeq 0$ rolls down to the minimum of the
potential.}
 \label{pVE3D}
\end{figure}

\subsection{Slow roll approximation}

 With Eq.~(\ref{Potential VE Hom}) in mind,
the field equation of motion and the Einstein equation in slow roll
approximation ($\ddot\psi_E\ll 3 H_E\dot\psi_E$, $\dot\psi_E^2\ll
V(\psi_E)$) become,
\bea
3H_E\dot\psi_E&=&-V_E^\prime(\psi_E)\,,\label{FrEqEinFr1}\\
H_E^2&=&\frac{8\pi G_N}{3}\, V_E(\psi_E)\,,\label{FrEqEinFr2}\\
\dot H_E&=&-\frac{\dot\psi_E^2}{2 M_{\rm P}^2}\,,
 \eea
where $H_E=\dot a_E/a_E$ and the metric tensor is
$g_{E\mu\nu}=\mbox{diag}[-1,a_E^2(t),a_E^2(t),a_E^2(t)]$.\\
The slow roll parameters are,
\bea &&\epsilon_E(\psi)=-\frac{\dot H_E}{H_E^2}\simeq\frac{2{F^\prime}^2}{F+\frac32{F^\prime}^2}\,,\label{epsilon}\\
&&\eta_E(\psi)=\frac{\dot \epsilon_E}{\epsilon_E H_E}\simeq
\frac{2F(2F{F^{\prime\prime}-{F^\prime}^2)}}{\big(F+\frac32{F^\prime}^2\big)^2}\,,\label{eta}\\
&&\xi_E(\psi)=\frac{\dot \eta_E}{\eta_E H_E}
 \label{xi}\\
&\simeq& \frac{2F^\prime}{F+\frac32{F^\prime}^2}
\Bigg(\frac{2F^2F^{\prime\prime}}{2FF^{\prime\prime}-{F^\prime}^2}
                    -
                    \frac{F^\prime\big(F-\frac32{F^\prime}^2+6F^{\prime\prime}\big)}{F+\frac32{F^\prime}^2}\Bigg)
\nonumber
\,,
\eea
where $F^\prime=dF(\psi)/d\psi$.
The spectra of scalar and tensor perturbations are,
\begin{eqnarray}
\Delta_s^2(k)&=& \Delta_s^2(k_*)\bigg(\frac{k}{k_*}\bigg)^{n_s-1}
,\quad   \Delta_s^2(k_*) = \frac{H_{E}^2}{8\pi^2\epsilon_{E}M_{\rm
P}^2}\,,
\nonumber\\
 \Delta_t^2(k)&=& \Delta_t^2(k_*)\bigg(\frac{k}{k_*}\bigg)^{n_t}
\,,\quad   \Delta_t^2(k_*) = \frac{2H_{E}^2}{\pi^2M_{\rm P}^2} \,,
\label{two spectra}
\end{eqnarray}
where $H_E$ and $\epsilon_E$ are to be calculated at the first Hubble crossing
during inflation at $t=t_*$, which is defined by $k=k_*=H_E(t_*)a_E(t_*)$.
The spectral indices $n_s$ and $n_t$ can be determined from the
variation of $\Delta_s^2(k)$  and  $\Delta_t^2(k)$ with respect to
$k$ at the first Hubble crossing. At the leading order in slow roll
approximation this procedure gives,
\begin{eqnarray}
  n_s &=& 1 + \bigg(\frac{d\ln[\Delta_s^2(k)]}{d\ln(k)}\bigg)_{k=k_*}  \simeq -2\epsilon_E-\eta_E
\,,
\label{ns}\\
n_t &=& \bigg( \frac{d\ln[\Delta_t^2(k)]}{d\ln(k)}\bigg)_{k=k_*}
\simeq -2\epsilon_E
\,,
 \label{nt}
\end{eqnarray}
where $k_*$ is a fiducial comoving momentum scale. To be in accordance with
Ref.~\cite{Ade:2015lrj} we choose $k_*=0.05~{\rm Mpc^{-1}}$.
From Eqs.~(\ref{two spectra}) it follows that the ratio of the
tensor and scalar spectra is,
\begin{equation}
   r(k_*)\equiv \frac{ \Delta_t^2(k_*)}{\Delta_s^2(k_*)} \simeq  16\epsilon_E
\,. \label{r}
\end{equation}
The running of the spectral index $n_s$ is,
\begin{equation}
 \alpha(k_*) = \bigg[\frac{d(n_s)}{d\ln(k)}\bigg]_{k=k_*} \simeq -(2\epsilon_E+\xi_E)\eta_E
\,. \label{running alpha}
\end{equation}
The number of e-folds can be calculated exactly for $F(\psi)$ given in Eq.~(\ref{F Psi}),
\begin{eqnarray}
 N(\psi) &=& \int_{t(\psi)}^{t(\psi_e)}H_E(\tilde t\,)d\tilde t
      \simeq \frac 12\int_{\psi}^{\psi_{e}}d\tilde\psi\bigg[\frac32\frac{F^\prime}{F}+\frac{1}{F^\prime}\bigg]
\nonumber\\
  &=& \frac34\ln\bigg(\frac{F(\tilde\psi)}{M_{\rm P}^2}\bigg)
      +\frac{1}{8\xi_2}\ln\bigg(\frac{M_{\rm P}^2F^\prime(\tilde\psi)}{\tilde\psi^3}\bigg)
              \bigg|_{\psi}^{\psi_{e}}\,,
\nonumber\\ \label{number e-folds}
\end{eqnarray}
where $\psi_e$ is the value of the field for which
$\epsilon_E=1$, at which point inflation ends. The number
of e-folds is defined to be zero at the end of inflation,
$N(\psi_e)=0$.

\subsection{Main Results}
\label{Main Results}

Since in this paper we analyze inflation in the center of a large
global monopole (for which the gradient terms can be neglected),
the potential in Jordan frame can be approximated by a constant
value, $V(\phi)\simeq \lambda \phi_0^4/4 = \mbox{const}$.
Since topological inflation requires a super-Planckian $\phi_0$,
the COBE normalization of the scalar spectrum in~(\ref{two spectra}),
 $ \Delta_s^2(k_*) \simeq 2.2\times 10^{-9}$~\cite{Ade:2015lrj}
measured at the pivotal comoving scale $k_*=0.05~{\rm Mpc}^{-1}$
implies, $\lambda\simeq 10^{-9}[10^2\epsilon_E](2M_{\rm P}/\phi_0)^4$,
representing a moderate fine tuning of the potential, which we do not address
any further in this work.
For the inflationary model with a constant potential and nonminimally
coupled scalar field given by Eqs.~(\ref{ActionPsi}--\ref{F Psi}),
we have shown~\cite{Glavan:2015aqa} that the spectral index $n_s$
exhibits a strong dependence on $\xi_2$ and a weak dependence on $\xi_4$
and  for $N(t_*)\simeq 62$ peaks at about,
\beq
n_s\simeq 0.955,
\quad \mbox{for}\quad \xi_2 \simeq -0.002
\quad \mbox{and} \quad \xi_4 \simeq -0.1
\,.
\label{ns xi3 xi4}
\eeq
In Ref.~\cite{Glavan:2015ora} we have shown that the upper bound on $n_s$ can be increased by,
for example, increasing $N(t_*)$, which can be done by constructing models in which the average
principal slow roll parameter during inflation is larger or/and by resorting to
non-conventional models in which there is a post-inflationary period of kination~\cite{Joyce:1997fc}.
The tensor-to-scalar ratio $r$ is generically small in this model
and rather strongly depends on $\xi_4$,
\beq
r\simeq \frac{10^{-6}}{|\xi_4|}
\,,
\label{r vs xi4}
\eeq
from which it follows that for $r$ of the order $10^{-2}$, $|\xi_4|$
has to be of the order $10^{-4}$. However, the price to pay is that
such a small $\xi_4$ reduces $n_s$, thus moving it away from the Planck sweet spot.\\
 The running of the spectral index
$\alpha$ in this model is negative and its magnitude is of the order of $10^{-3}$.
The recent Planck Collaboration analysis~\cite{Ade:2015lrj}
gives for the scalar spectral index, $n_s=0.968\pm 0.006$ ($1\sigma$ error bars)
which is obtained by fixing $\alpha=0$. However, relaxing that constraint
reveals that there is slight preference for a negative running,
$\alpha= -0.003\pm 0.007$, whereby
the central value for $n_s$ decreases and the error bars
increase somewhat to, $n_s = 0.965\pm0.010$ (see figure~3 of Ref.~\cite{Ade:2015lrj}).
Finally, $r_{0.05}<0.12$. Taking these latter values
as more realistic, we conclude that
the present model is in good agreement (at the
$1\sigma$ level) with the currently available CMB and LSS data~\footnote{When this work was nearing
completion, a new article appeared~\cite{Palanque-Delabrouille:2015pga} in which
the most recent Ly$\alpha$ data have been analyzed. The new measurements have further constrained $n_s$ and
$\alpha$, such that the new most favorite values are
$n_s=0.963\pm0.045$ and $\alpha=-0.0104\pm0.0031$ ($1\sigma$ error bars).
Our model lies about 2 standard deviations from these values (mostly because it predicts
a too small $n_s$ and also prefers a rather small $|\alpha|$.
Most of other single field inflationary models also lie at least two standard deviations from the new sweet spot
of $n_s$ and $\alpha$.
Because in our model $\eta_E$ is rather large, the resulting  $|\alpha|$ is also rather large, but
still not large enough to agree better than two sigmas with the results of Ref.~\cite{Palanque-Delabrouille:2015pga}.
If these results get confirmed by independent measurements, they will severely constrain many (single field) inflationary models.
} .

 The main obstacle for getting an even better agreement with the data is
 an upper limit on $n_s=-2\epsilon_E-\eta_E$ that we see from numerical considerations.
 Since $r=16 \epsilon_E\sim 10^{-6}/|\xi_4|$ can be tiny,
the limitation must come from $\eta_E$, {\it i.e.} we have
$\eta_E>0.045$, which we refer to the {\it moderate $\eta$ problem}
(as opposed to the traditional $\eta$ problem~\cite{Easson:2009kk},
where $\eta$ is induced by higher dimensional operators and
can be as large as the order unity).
It would be of interest to investigate whether similar cures
tried with the traditional $\eta$ problem help to alleviate our moderate $\eta$ problem.
One such cure could be the effect of decays of $\psi$ during inflation~\cite{Berera:2004vm}.

In Refs.\cite{Notari1,Notari2}, the authors investigated a model of
inflation from a false vacuum in which the exit from inflation is
obtained by tunneling to a true vacuum in virtue of the nonminimally
coupled scalar field of the form similar to our model, Eq.~(\ref{F
Psi}). Apart from having different mechanisms for ending inflation,
the main difference between theirs and our model is that they
analyze two different regimes, for small and large fields, in which,
first quadratic and then quartic coupling dominates, respectively.
Here we have shown that both couplings are relevant and non-zero in
both regimes, and only as such lead to the spectral index $n_s$ and
tensor-to-scalar ratio $r$ that are in good agreement with the
currently available CMB and LSS data.


\section{Generic initial conditions}
\label{Generic initial conditions}

 Broadly speaking, there are two classes of generic initial conditions.
The universe may have started in a very energetic state,
with an energy density and pressure close to the Planck density,
$\rho\sim p\sim M_{\rm P}^4$. Alternatively, the initial Universe
may have been an (almost) empty state, whose energy density is
dominated by the potential energy of (one or many) scalar fields.
Broadly speaking, the former initial conditions are "chaotic"~\cite{Linde:1983gd},
and prominently figure in chaotic inflationary models, bounce models, {\it etc.}
Here we shall refer to this class of initial conditions as {\it stochastic initial conditions.}
A prominent example of the latter are landscape models, which generically yield eternal inflation,
and which are supported by the semiclassical approaches
to the Universe's creation~\cite{Vilenkin:1984wp,Hartle:1983ai}.

A typical state of quantum fields in a {\it stochastic initial state} are wildly fluctuating quantum fields, whose
amplitude can be as large as the Planck scale (but not much larger,
as that would cost super-Planckian amount of gradient energy),
$\psi_i\sim M_{\rm P}$,
and whose total energy density and pressure are Planckian, such that quantum gravitational effects are
large, and therefore we can say nothing reliable about the evolution of the Universe from such a state.
After the Universe expands somewhat, thereby cooling to a sub-Planckian density, we have,
$\rho_{\rm in}\sim p_{\rm in} \ll M_{\rm P}^4$ and perturbative treatments apply.
If there is no large (Planckian) mass scale in the problem, every Hubble region of
this universe will evolve to a good approximation as,
\begin{equation}
\langle H^2\rangle_V\sim \frac{\langle \rho\rangle_V}{M_{\rm P}^2} \propto \frac{1}{a^4}
\,,
\nonumber
\end{equation}
where the averaging $\langle\cdot\rangle_V \equiv {\rm Tr}[\hat \rho_V\; \cdot]$
($\hat \rho_V$ is a suitably coarse grained density operator)
is assumed to be taken over a volume $V\gtrsim 3\pi/[3H^{3}]$ that is larger than, but comparable to, the Hubble volume.
In other words, the very early Universe expands as radiation dominated. On the other hand, quantum
fields scale approximately conformally:
scalars scale as $\psi_i\propto 1/a$ and fermionic fields scale as $\psi_j\propto 1/a^{3/2}$, {\it etc.}
From the Planck density $\rho\sim M_{\rm P}^4$ to the GUT density $\rho\sim E_{\rm GUT}^4$ ($E_{\rm GUT}\sim 10^{16}~{\rm GeV}$,
the scale factor expands by about, $a_{\rm GUT}/a_{\rm Planck}\sim 10^3$.
In our model, the GUT scale is the scale when the symmetry $O(3)$ gets broken to $O(2)$
 and global monopoles form by the Kibble mechanism, triggering inflation.
From the Planck regime to the GUT regime the amplitude of scalar (fermionic) field fluctuations
decreases by a factor of $\sim 10^3$ ($\sim 3\times 10^{4}$), implying that the amplitude of
typical scalar field fluctuations is $\psi_i\sim 10^{-3}~M_{\rm P}$, which is much larger than
the amplitude of fluctuations in a Bunch-Davies vacuum,
$(\psi_i)_{\rm BD}\sim H_{\rm GUT}\sim 10^{-6}~M_{\rm P}$. From this we conclude that
the amplitude of scalar field fluctuation at the GUT scale is most likely in the range,
\begin{equation}
10^{-6}~M_{\rm P} \lesssim   \psi_i \lesssim 10^{-3}~M_{\rm P}
\,.
\label{typical amplitude}
\end{equation}

An important question is whether the range of initial field values in Eq.~(\ref{typical amplitude})
is consistent with the requirement
on the minimum number of e-folds of inflation, $N_0>65$ in our model.

\begin{figure}
\centering
\includegraphics[scale=0.6]{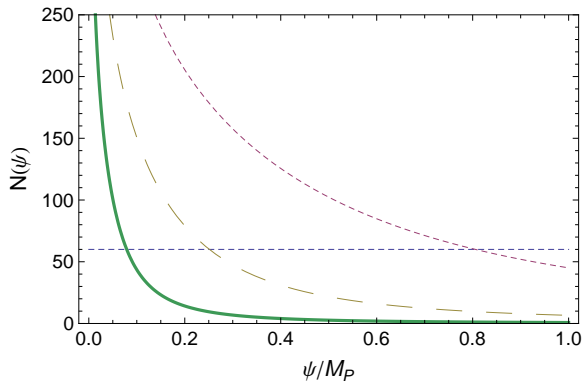}
\caption{The number of e-folds as a function of $\psi$ for
$\xi_2=-0.002$ and different values of $\xi_4$: $\xi_4=-0.001$ (short dashed),
$\xi_4=-0.01$ (long dashed) and $\xi_4=-0.1$
(thick solid). All three curves show that $N_0\gg 60$
when the initial field amplitude is in the range of Eq.~(\ref{typical amplitude}).}
 \label{pN0a}
\end{figure}

\begin{figure}
\centering
\includegraphics[scale=0.65]{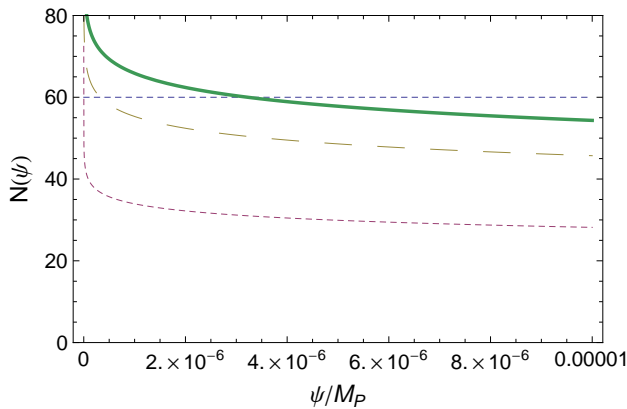}
\caption{The number of e-folds as a function of $\psi$ for $\xi_4=-0.1$
and different values of $\xi_2$: $\xi_2=-0.1$ (short dashed),
$\xi_2=-0.06$ (long dashed) and $\xi_2=-0.05$ (thick solid).
Here $N_0(10^{-6}~M_{\rm P})>60$ for $|\xi_2|\lesssim 0.06$.}
 \label{pN0b}
\end{figure}

 To address this question, in figures~\ref{pN0a} and~\ref{pN0b} we show how the number of e-folds of inflation
depends on the initial field value and on the two nonminimal
couplings, $\xi_2$ and $\xi_4$.
From figure~\ref{pN0a} we see that for a smallish value $\xi_2=-0.002$ favoured by the CMB data~(\ref{ns xi3 xi4}),
for all initial values of the field $\psi$ in the interval~(\ref{typical amplitude})
and independently on the value of $\xi_4$ (as long as it is in the interval $-0.001>\xi_4>-0.1$),
one generically gets a much larger number of e-folds than what is required, {\it i.e.} $N_0\gg 60-70$.
Note that the number of e-folds falls quite dramatically as $|\xi_4|$ increases.

To study that effect in more detail, in figure~\ref{pN0b} we show how the number of e-folds depends on $\xi_2$ for
a fixed value of $\xi_4$. For definiteness, we chose a rather large value, $\xi_4=-0.1$, see Eq.~(\ref{ns xi3 xi4}).
Similarly as in figure~\ref{pN0a} we see that the number of e-folds of inflation decreases quite dramatically
as $|\xi_2|$ increases. What is important to note is that, close to the preferred value of the parameters, where
$\xi_2$ is rather small ($\xi_2\sim -10^{-3}$ ) and $\xi_4$ is not too large,
one gets the number of e-folds, $N_0\gg 60-70$. In particular, to get
an $r$ that is observable by the near future efforts ($r\gtrsim 10^{-2}$), $|\xi_4|$ must be quite small.
Eq.~(\ref{r vs xi4}) implies $\xi_4\simeq -10^{-6}/r$, and thus imposing $r\gg 10^{-3}$ yields $|\xi_4|\ll 10^{-3}$,
for which $N_0\gg 60-70$.

To complete the discussion on how generic the initial conditions that lead to inflation are, we also need to estimate the
likelihood that an observer will experience at least 60 e-folds of inflation. Two relevant probabilities can be defined:
the {\it a priori} probability, $P_{\it a priori}$, which can be defined as the fraction of the initial volume of the Universe
that exhibits at least $60$ e-folds of inflation,
\begin{equation}
 P_{ a\; priori} = \frac{V(r_c)}{V_M}
\label{a priori probability}
\end{equation}
where $V(r_c)$ is the volume of the monopole core (which has radius $r_c$) that exhibits $60$ or more e-folds of
inflation, and $V_M=(4\pi R_H^3/3)/p_M$ is the average volume occupied by one monopole (if the monopoles form
at a phase transition by the Kibble mechanism then $p_M\simeq 1/8$ or larger~\cite{Prokopec:1991ab}) and $R_H=1/H$ is
the Hubble radius at the time of monopole formation. On the other hand, after inflation different parts of the Universe have
expanded by different amounts, and so after inflation one can define {\it a posteriori} probability that an observer will be
in a universe that has inflated at least $60$ e-folds as,
\begin{eqnarray}
 P_{a\; posteriori} &=& \frac{ \langle{\rm e}^{N_{\rm interior}}\rangle V(r_c)}
               { \langle{\rm e}^{N_{\rm interior}}\rangle V(r_c) + \langle{\rm e}^{N_{\rm exterior}}\rangle V_M}
\nonumber\\
   &=& \frac{P_{a\; priori} }{P_{a\; priori}
  +\frac{ \langle{\rm e}^{N_{\rm exterior}}\rangle }{ \langle{\rm e}^{N_{\rm interior}}\rangle }}
\,,
\label{a posteriori probability}
\end{eqnarray}
where $\langle{\rm e}^{N_{\rm interior}}\rangle$ ($\langle{\rm e}^{N_{\rm exterior}}\rangle$) denotes the average
number of e-folds in the interior (exterior) region, in which the number of e-folds is above (below) $60$.
We shall not go into the subtlety related to defining the average number of e-folds, as the proper definition
requires knowledge of what the right (volume) measure is. We just note here that there is no agreement in literature
(on eternal inflation) on how to define the measure. Barring that difficulty, Eq.~(\ref{a posteriori probability})
implies that, when
\begin{equation}
\langle {\rm e}^{N_{\rm exterior}}\rangle / \langle {\rm e}^{N_{\rm interior}}\rangle \ll P_{a\; priori},
\label{condition for high prob}
\end{equation}
then
\begin{equation}
 P_{a\; posteriori} \approx 1
\,.
\label{a posteriori probability:2}
\end{equation}
When this condition is met then most of the late time observers
will find themselves in a universe which inflated (more than 60 e-folds) in the past (light-cone).
In order to get an idea how likely the condition~(\ref{condition for high prob}) is satisfied,
we shall now estimate $P_{a\; priori}$. From numerical simulations~\cite{Marunovic:2013eka},
\cite{Vilenkin:1994pv} it is known that
the field $\phi_E$ in the monopole core grows linearly with the radius, $\phi_E\simeq \beta r$,
where $\beta \sim \Delta\phi_E/\Delta r\sim \phi_0/[\delta_0\sqrt{F(\psi)}]\sim \sqrt{\lambda}\phi_0^2$.
The field grows with the distance from the center of the monopole and reaches a critical value $\phi_c$ at a
(comoving) distance $r_c$,
$\phi_c\sim \sqrt{\lambda}\phi_0^2 r_c$. Inserting $V(r_c)=(4\pi/3)r_c^3$ into~(\ref{a priori probability})
one obtains,
\begin{equation}
 P_{ a\; priori} \sim \frac{p_M}{\sqrt{27}}\left(\frac{\phi_c}{M_P}\right)^3
\,,
\label{a priori probability:2}
\end{equation}
where $\phi_c$ is the critical field value that can be estimated as follows.
From Eq.~(\ref{phiI})  we then infer that, for $N\simeq 62$ and for a typical choice of the couplings,
$\xi_2=-0.001$, $\xi_4=-0.1$, $\phi_0=2.4\sqrt{3}M_{\rm P}$, one gets
$\phi_c\sim 10^{-6}~{\rm M_{\rm P}}$. When this is inserted into~(\ref{a priori probability:2}) one obtains,
$ P_{ a\; priori} \sim 2\times 10^{-20}$, which means that -- in order for (\ref{a posteriori probability:2}) to be satisfied --
one needs the average number of e-folds in inflating patches (with $N>60$) to be by at least by $\sim 45$ larger than
the average number of e-folds in the regions of the Universe with no or little inflation. Even though, based on the
above analysis we cannot claim that this is indeed achieved, given that the center of the monopole exhibits eternal inflation,
it seems reasonable to posit that the condition~(\ref{condition for high prob}) will be quite generically met.
A more comprehensive numerical study that includes spatial inhomogeneities is required to fully answer
this intriguing question.

\begin{figure}
\centering
\includegraphics[scale=0.6]{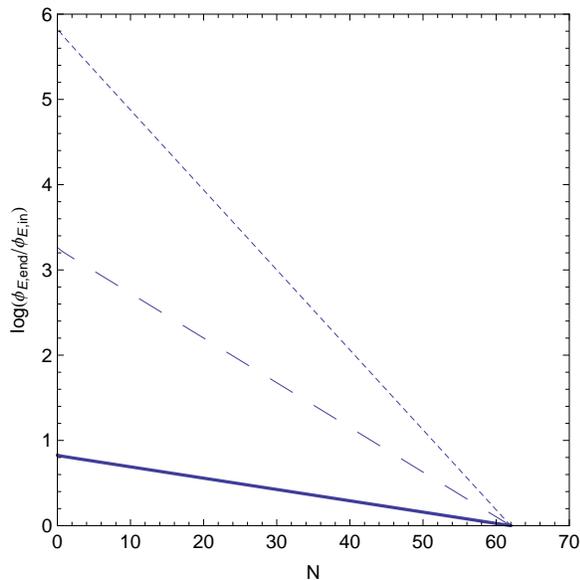}
\caption{$\log(\phi_{E,end}/\phi_{E,in})$ as a function of the
number of e-folds ($N = 0$ denotes the end of inflation) for
$\delta_0 = 2.4 H_0^{-1}$ (short dashed curve), $\delta_0 = 5
H_0^{-1}$ (long dashed curve) and $\delta_0 = 10 H_0^{-1}$ (solid
curve). $\xi_2 = -0.001$, $\xi_4 = -0.1$.}
 \label{pAmplPhi}
\end{figure}


\section{The fate of the monopole}
\label{The fate of the monopole} \vskip -0.1cm

 This section is fully devoted to a discussion of the graceful exit problem,
 {\it i.e.} on how inflation ends in our model. The insights we attain
constitute the principal results of this paper. We shall begin by
presenting a simple model for the expansion of the monopole core
during inflation. This is followed by a simple calculation of the
core evolution in the postinflationary epochs. We first discuss the
$\epsilon_E=4/3$ epoch, and then the subsequent radiation era.

If the monopole has a super-Hubble core during inflation, topology
cannot prevent it from expanding. This can be seen intuitively by
observing that on super-Hubble scales the space expands
super-luminally and hence nothing can prevent expansion of the
monopole core. To estimate the rate of the core expansion, we
linearize the monopole equation and neglect all gradients (which is
justified on super-Hubble scales, on which $\|\nabla/a\|\ll H$) to
get,
\begin{equation}
\Big(\frac{d^2}{dt^2}+3H_E \frac{d}{dt} -\mu_E^2\Big)\phi_E(t)\simeq
0 \,, \label{eom large core}
\end{equation}
where
\beq
-\mu_E^2=\frac{dV_E^2}{d\phi_E^2}|_{\phi_E=0}=-\frac{\lambda\phi_0^2}{F(\psi)/M_{\rm P}^2}\equiv
-\frac{\mu_0^2}{F(\psi)/M_{\rm P}^2} \eeq
 is the curvature of the
monopole potential at the origin and $\mu_0^2=\lambda\phi_0^2$.
Friedmann equation~(\ref{FrEqEinFr2}) allow us to express $\mu_E$ in
terms of $H_E$:
\beq \mu_E^2=\frac{\mu_0^2}{H_0}H_E\,,\qquad H_0^2=\frac{8\pi
G_N}{3}\frac{\lambda\phi_0^4}{4}\,.
\eeq
Now, the equation of motion~(\ref{eom large core}) in all epochs can
be written in terms of the Hubble parameter as,
\begin{equation}
\Big(\frac{d^2}{dt^2}+3H_E \frac{d}{dt}
-\frac{\mu_0^2}{H_0}H_E\Big)\phi_E(t)\simeq 0 \,, \label{eom large
core 2}
\end{equation}

{\bf A. The $\epsilon_E\simeq 0$ inflationary epoch.}
 Approximating inflation by de Sitter space,
$H_E\simeq {\rm constant}$, and  making the {\it Ansatz},
$\phi_E\propto {\rm e}^{\lambda t}$, we get a quadratic equation for
the root $\lambda$,
\begin{equation}
   \lambda^2 + 3H_E\lambda -\frac{\mu_0^2}{H_0}H_E=0
   \,.
\label{two roots}
\end{equation}
The positive solution
\begin{equation}
   \lambda=\lambda_+=\frac{3H_E}{2}\Bigg[\sqrt{1+\frac 49 \frac{\mu_0^2}{H_0H_E}}-1\Bigg]
      \label{positive lambda}
\end{equation}
is the relevant solution, since it gives the growing solution for
$\phi_E$ that eventually dominates. Since $H_E$ changes slowly in
time during most of inflation (in the sense that $|\dot H_E|\ll
H_E^2$), we can utilize adiabatic approximation and write the
solution for $\phi_E$ as an integral $\phi_E(t)\sim \phi_{E\rm in}
\exp{\int\lambda_+ dt}$, which in slow-roll approximation becomes an
integral over $\bar\psi$ ($\bar\psi=\psi/M_{\rm P}$, $\bar
F=F/M_{\rm P}^2$),
\beq
\frac{\phi_E(t)}{\phi_{E\rm in}}\sim  \exp{\Bigg\{\! \frac 34 \int d\bar\psi\bigg(\!\frac{1}{\bar F'}\!+\!\frac{3\bar F'}{2\bar F}\bigg)
\!\left(\sqrt{1\!+\!\frac 49\frac{\mu_0^2}{H_0^2}\bar
F}\!-\!1\right)\!\Bigg\}}\,.
\label{phiI}
\eeq
During inflation the scale factor grows as, $a_E\propto \exp{(\int dt H_E )}$, so the
dynamical monopole core size in an accelerating universe, $R_c(t)\propto a_E(t)/\phi_E(t)$~\cite{Vilenkin:1994pv} compared with the
Hubble radius $R_H=H_E^{-1}$ is given by,
\begin{eqnarray}
 \frac{R_c(t)}{R_H}&\propto& \frac{H_E}{\phi_{E\rm in}}\exp \Bigg\{\frac 34 \int
d\bar\psi\bigg(\!\frac{1}{\bar F'}\!+\!\frac{3\bar F'}{2\bar F}\bigg)
\nonumber\\
&&\hskip 1.5cm
\times \left(\!\frac{5}{3}\!-\!\sqrt{1\!+\!\frac 49\frac{\mu_0^2}{H_0^2}\bar
F}\right)\!\Bigg\}
\,.
\label{RcI}
\end{eqnarray}
A numerical solution of  this equation shows that during inflation
the monopole core size expands exponentially with respect to the Hubble radius
as long as $\mu_0^2\bar F/H_0^2\ll 1$. If during inflation the term in the second line of Eq.~(\ref{RcI})
become negative, the monopole core will start shrinking during inflation.
The critical value of the field when that happens is,
\begin{equation}
  \bar F_{\rm cr}=\frac{F(\psi_{\rm cr})}{M_{\rm Pl}^2} = \frac{4H_0^2}{\mu_0^2}
\,.
\label{Fcr}
\end{equation}
If this value is reached before the end of inflation ({\it i.e.} before $\epsilon_E$ reaches unity),
then the monopole will start shrinking.

 The fate of the monopole
after inflation depends on whether it shrinks fast enough
 to avoid~\footnote{Fragmentation is the process by which one large monopole breaks up into smaller pieces
due to the instability to growth of small scale perturbations. To accurately model this process one would have to
include spatial gradients and perfore a full three dimensional numerical evolution,
which is beyond the scope of this paper.}
to smaller pieces. A detailed investigation of this interesting question
we leave for future work.
In the coming subsections we discuss the post-inflationary monopole dynamics in some detail.
We separately consider three cases: (1) the monopole enters a prolonged $\epsilon_E=4/3$ epoch;
(2) the fields decay quickly after inflation such that almost instantly one enters a radiation era, and (3) the hybrid case,
in which the monopole enters a short $\epsilon_E=4/3$ period, which is followed by radiation.
For a somewhat detailed discussion on how fields may decay after inflation we refer to~\cite{Glavan:2015ora}.

{\bf B. The $\epsilon_E=4/3$ epoch.}
Let us now solve Eq.~(\ref{eom large core 2}) for epoch in which
$\epsilon_E=4/3$ by using $H_E=1/(\epsilon_E t)$,
\begin{equation}
\Big(\frac{d^2}{dt^2}+\frac{3}{\epsilon_E
t}\frac{d}{dt}-\frac{\mu_0^2}{H_0\epsilon_E}\frac 1t\Big)\phi_E(t) =
0 \,.
 \label{eom large core:conformal:3}
\end{equation}
The two linearly independent (real) solutions of Eq.~(\ref{eom large
core:conformal:3}) are the two modified Bessel functions of the
order $5/4$,
\begin{equation}
 \phi_E(t)=\phi_E(y)=\alpha y^{-5/4}I_{5/4}\big(y\big)
  +\beta
  y^{-5/4}I_{-5/4}\big(y\big)
  \,,
\label{phiE4/3}
\end{equation}
where
\beq y=\frac{3\mu_0}{2H_0}\bar F(\psi)^{1/2}\,. \eeq
Constants $\alpha$ and $\beta$ are obtained by the smooth matching
of the solutions at the end of inflation/beginning of the
$\epsilon_E=4/3$ epoch. Relative to the Hubble radius,
$R_H=(H_0/\bar F)^{-1}$, the monopole core scales as
\beq
  \frac{R_c}{R_H}\propto \frac{\bar F^{-1+1/\epsilon_E}}{\phi_E} \,.
\label{monopole core in 43}
\eeq
\begin{figure}
\centering
\includegraphics[scale=0.65]{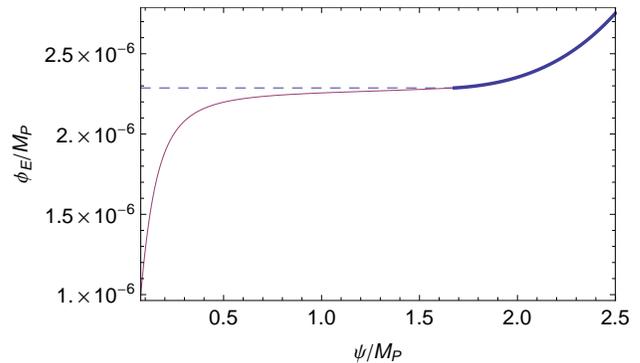}
\caption{The monopole field $\phi_E/M_{\rm P}$  as a function of
 $\psi/M_{\rm P}$ grows during inflationary era (thin curve) and smoothly matches the field in
  $\epsilon_E=4/3$ epoch (thick curve).  ($\xi_2=-0.001$, $\xi_4=-0.1$, $\delta_0=10 H_0^{-1}$,
$\phi_0=10\sqrt{3}M_{\rm P}$,
 $\phi_{E\rm in}=10^{-6}~M_{\rm P}$, $N\simeq 62$.)}
 \label{pPhiE}
\end{figure}
From Fig.~\ref{pPhiE} we see that the monopole field during
inflationary era indeed grows in time and smoothly matches the field
in $\epsilon_E=4/3$ era, during which the monopole starts shrinking dramatically after $F$ reaches its critical
value, $y_{\rm cr}\sim 1$, which is approximately given by the inflationary critical value~(\ref{Fcr}).

 Two scenarios are therefore envisaged. In the first {\it subcritical} scenario the monopoles
effective mass remains small when compared with the
Hubble radius both during inflation and the subsequent $\epsilon_E=4/3$ epoch. In this case the small argument
expansion of  the solution~(\ref{phiE4/3}) applies, $ \phi_E(t)\propto y^{-5/2}\propto \bar F^{-5/4}$, such that
Eq.~(\ref{monopole core in 43}) reduces to,
\beq
  \frac{R_c}{R_H}\propto \bar F\propto
\exp\bigg(\frac43N_{4/3}\bigg) \,,
\label{monopole core in 43:b}
\eeq
implying that the monopole continues expanding.
Such a monopole can decay only by the process of fragmentation,
according to which it will break into smaller sub-Hubble size pieces (which may, but need not be, topologically charged).
The charged pieces will subsequently shrink to their Minkowski size or mutually annihilate if they are oppositely charged.
We shall not discuss this case any further.

  Here we are more interested in the case when the monopole reaches its critical size during inflation or during
the $\epsilon_E=4/3$ epoch. In the latter case, $y\gg 1$ and the large $y$-expansion of the solution~(\ref{phiE4/3})
applies, $\phi_E\propto \exp(y)/y^{7/4}$. In this case Eq.~(\ref{monopole core in 43}) becomes,
\begin{eqnarray}
  \frac{R_c}{R_H}&\propto& \bar F^{5/8}\exp\bigg(\!\!-\!\frac{3\mu_0}{2H_0}\sqrt{\bar F(\psi)}\bigg)
\nonumber\\
&\propto& \exp\bigg(\!\!-\!\frac{3\mu_0}{2H_0}\sqrt{\bar F(\psi_e)}{\rm e}^{2N_{4/3}/3}\bigg)
 \,,
\label{monopole core in 43:c}
\end{eqnarray}
where $N_{4/3}$ denotes the number of e-folds during the
$\epsilon=4/3$ epoch and in the last term we neglected the prefactor
$\bar F^{5/8}$. If the prefactor $\frac{3\mu_0}{2H_0}\sqrt{\bar
F(\psi_e)}$  in the exponent in~(\ref{monopole core in 43:c}) is
larger than one, then the monopole starts shrinking already during
inflation. From this result we see that the monopole whose effective
mass becomes larger than the Hubble radius shrinks dramatically.
Indeed, take the simple example in which the monopole expands during
inflation by a factor $R_c/R_H\sim\exp(N_0)$, where $N_0$ is the
total number of e-folds (a typical number of e-folds in our model is
100s or 1000s but not much larger than that) and it attains its
critical size~(\ref{Fcr}) at the end of inflation, {\it i.e.}
$\sqrt{\bar F(\psi_e)} =2H_0/\mu_0$.
 In that case from Eq.~(\ref{monopole core in 43:c}) we get,
\begin{eqnarray}
N_{4/3}\simeq \frac32\ln\left(\frac{N_0}{3}\right)
 \,,
\label{monopole core in 43:d}
\end{eqnarray}
which for $N_0=100$ evaluates to $N_{4/3}\simeq 5.4$ and for $N_0=1000$, $N_{4/3}\simeq 8.7$,
implying that the monopole shrinks to the Hubble size within a few e-folds. After that --
thanks to the workings of the gradient terms -- the monopole shrinks
rapidly to its flat (Minkowski) space size.

\begin{figure}
\centering
\includegraphics[scale=0.65]{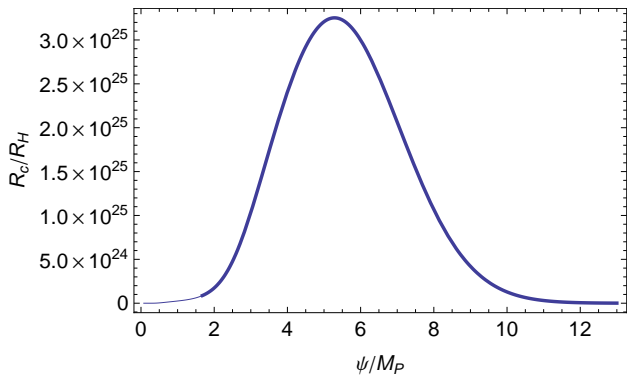}
\caption{During inflation, the monopole core radius compared to the
Hubble radius $R_c/R_H$,  grows in $\psi/M_{\rm P}$ (thin curve),
smoothly matches onto its value in $\epsilon_E=4/3$ era, and after
some time rapidly shrinks to its Minkowski size (thick curve).
($\xi_2=-0.001$, $\xi_4=-0.1$, $\delta_0=10 H_0^{-1}$,
$\phi_0=10\sqrt{3}M_{\rm P}$,
 $\phi_{E\rm in}=10^{-6}~M_{\rm P}$, $N\simeq 62$.)}
 \label{pRc}
\end{figure}

\begin{figure}
\centering
\includegraphics[scale=0.65]{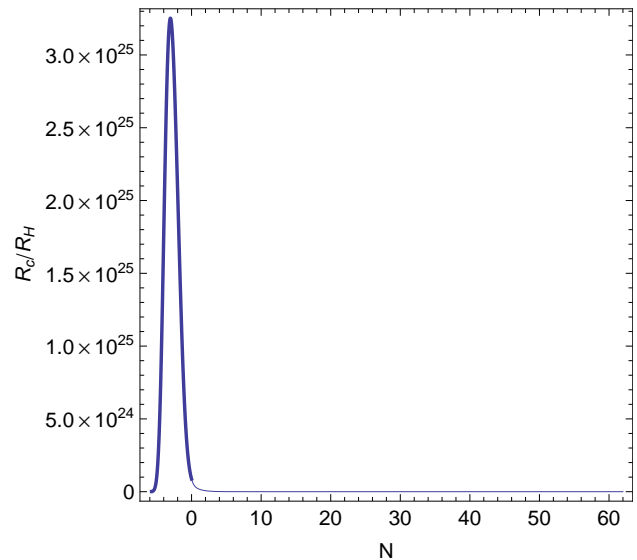}
\caption{The monopole core radius compared to the Hubble radius
$R_c/R_H$ as a function of the number of e-folds. During inflation
the monopole grows (thin curve), smoothly matches onto its value in
$\epsilon_E=4/3$ era (thick curve), and after $\Delta N~5$ e-folds
rapidly shrinks to its Minkowski size (thick curve).
($\xi_2=-0.001$, $\xi_4=-0.1$, $\delta_0=10 H_0^{-1}$,
$\phi_0=10\sqrt{3}M_{\rm P}$,
 $\phi_{E\rm in}=10^{-6}~M_{\rm P}$, $N\simeq 62$.)}
 \label{pRcB}
\end{figure}

The simple estimate~(\ref{monopole core in 43:d}) agrees quite well
with numerical calculations, example of which is shown in
figures~\ref{pPhiE} and~\ref{pRc}. The numerical example in
Fig.~\ref{pRc} illustrates well the simple analytic estimates
presented above. In particular, figure~\ref{pRc} shows the case in
which the monopole core size during inflation grows, continues to
grow for a while during the $\epsilon_E=4/3$ epoch (since it still
has not attained its critical size), but it eventually begins to
shrink rapidly to the size comparable to the Hubble radius and --
thanks to the gradient terms -- shrinks eventually to its flat space
size. For the given set of parameters, $\xi_2=-0.001$, $\xi_4=-0.1$,
$\delta_0= 10 H_0^{-1}$, $\phi_0=10\sqrt{3}M_{\rm P}$,
$\psi_0=10^{-6}~M_{\rm P}$ after $N_{4/3}\simeq 5-6$ e-folds  the
monopole will become sub-Hubble (see Fig.~\ref{pRcB}). This rapid
monopole shrinking will accumulate a lot of kinetic and gradient
energy, which will eventually get released into the production of
particles to which $\psi$ and $\phi^a$ couple (that includes some
gravitational wave production). That also means that the
$\epsilon_E=4/3$ epoch will be super-seeded by a radiation epoch.

{\bf C. The $\epsilon_E=2$ (radiation) epoch.} If $\psi$ decays rather
quickly after inflation, such that the $\epsilon_E=4/3$ epoch plays
no significant role for the monopole core dynamics, one can
approximate the transition from inflation to radiation era by a
sudden transition. In that case, the two linearly independent (real)
solutions of Eq.~(\ref{eom large core:conformal:3}) are
$\propto\exp(\pm z)/z$, where $z=2\sqrt{\mu_0^2 t/2H_0}$.
The analogous procedure as above tells us that, if the monopole is subcritical
during radiation then, $R_c/R_H\sim {\rm const.}$, {\it i.e.}
it expands as fast as the Hubble radius. Therefore, also in this case
the monopole starts shrinking only after it reaches its critical size.
The number of e-folds needed to reach the Hubble size is in this case,
\begin{equation}
 N_{\rm rad} \simeq \ln\Big(\frac{N_0}{3 [F(\psi_e)/F_{\rm cr}]^{1/2}}\Big)
\,. \label{monopole core in radiation}
\end{equation}
This implies that in radiation epoch a super-critical monopole shrinks even more rapidly than
in the $\epsilon_E=4/3$ epoch~(\ref{monopole core in 43:d}).

{\bf D. The $\epsilon_E=4/3$ epoch followed by radiation.}
 Finally, in the hybrid case, in which the $\epsilon_E=4/3$ epoch does not last long enough to shrink the monopole
to its sub-Hubble size, we can combine the two analyses from above to obtain
the number of e-folds needed for the monopole to become
sub-Hubble during radiation era. The result is,
\begin{equation}
N_{\rm rad}=-\frac{2}{3}N_{4/3}+\ln\left(\frac{N_0}{3
[F(\tilde\psi_e)/F_{\rm cr}]^{1/2}}\right) \,,
\end{equation}
where $\tilde\psi_e$ is the field value at the end of the $\epsilon_E=4/3$ epoch.
Of course, this formula is valid only when $N_{\rm rad}>0$.


\section{Discussion}
\label{Discussion}

 In this paper we present a new model of topological inflation with an additional
 nonminimally coupled scalar field. Our model produces cosmological perturbations
 with properties consistent with the existing CMB and LSS data. Furthermore,
thanks to the nonminimally coupled scalar field, graceful exit is
naturally
 realized in the model.

 The following interesting questions remain unanswered in this initial study of the model:

  \begin{itemize}
\item[(1)] Here we have considered the homogeneous case only which applies
in the limit when the monopole core $50-60$ e-folds from the end of inflation
is much larger than the Hubble radius. It would be of great interest to study
the case in which the effects due to the monopole spatial inhomogeneities are significant
and, if possible, connect them with the observed anomalies/anisotropies in the CMB observations.

\item[(2)] Our consideration of the exit from inflation is quite rudimentary.
One would like to refine it and study in detail some specific models
of preheating. In particular, we have found out that, quite
generically, inflation is followed by a period in which
$\epsilon_E=4/3$. It would be very important to investigate whether
any observational consequences of this epoch survive up to today. If
affirmative, if would be the smoking gun for this class of
inflationary models.

\item[(3)] Our numerical investigation shows that the spectral slope of scalar perturbations
$n_s$ cannot be larger than about $n_s=-2\epsilon_E-\eta_E\leq
0.955$, while the tensor-to-scalar ratio $r$ can be arbitrarily
small. This means that $\epsilon_E$ can be made as small as desired,
but that $\eta_E$ is limited from below by about $0.045$. This is a
weaker version of the well-known $\eta$-problem, which plagues many
large field inflationary models, since higher dimensional operators
that are argued to appear naturally in these models give $\eta\sim
1$. Various solutions have been proposed to the $\eta$-problem. It
would be of interest to check whether analogous solutions could be
used to relax the lower bound on $n_s$ that we found.

\item[(4)] Throughout the paper we have assumed that slow roll approximation
correctly characterizes cosmological perturbations. It would be of interest to study
the conditions under which slow roll approximation is correct, and when it fails.
In particular, it would be of interest to investigate whether one can get a better
agreement with the data by studying the influence of initial conditions that
require a treatment that goes beyond slow roll approximation.

\end{itemize}

\subsection*{Acknowledgments}
This work is part of the D-ITP consortium, a program of the
Netherlands Organization for Scientific Research (NWO) that is
funded by the Dutch Ministry of Education, Culture and Science
(OCW). AM is funded by NEWFELPRO,  an International Fellowship
Mobility Programme for Experienced Researchers in Croatia and by the
D-ITP.

\end{document}